\documentclass[11pt]{article}
\usepackage{graphicx}

\newcommand{\BABARPubYear}    {08}
\newcommand{\BABARProcNumber} {156}
\newcommand{\SLACPubNumber} {13509}

\long\def\inst#1{\par\nobreak\kern 4pt\nobreak
    {\it #1}\par\vskip 10pt plus 3pt minus 3pt}

\RequirePackage{xspace}

\def\Bbar    {\kern 0.18em\overline{\kern -0.18em B}{}\xspace}

\def\BBbar{\mbox{$B\overline {B}\ $}}
\def\Bz      {\ensuremath{B^0}\xspace}
\def\Bp      {\ensuremath{B^+}\xspace}

\def\Bzb     {\ensuremath{\Bbar^0}\xspace}
\def\K0S         {\ensuremath{K^0_S}\xspace}
\def\CP                {\ensuremath{C\!P}\xspace}
\def\ra                 {\ensuremath{\rightarrow}\xspace}

\def\calA{{\ensuremath{\cal A}}\xspace}
\def\babar{{\em B}{\footnotesize\em A}{\em B}{\footnotesize\em AR}}
\def\lbabar{\mbox{{\large\sl B}\hspace{-0.4em} {\normalsize\sl A}\hspace{-0.03em}{\large\sl B}\hspace{-0.4em} {\normalsize\sl A\hspace{-0.02em}R}}}

\def\calB{{\ensuremath{\cal B}}\xspace}
\def\RK{{\ensuremath{R_{K^{(*)}}}}\xspace}
\def\calAI{{\ensuremath{{\cal A}^{K^{(*)}}_I }  }\xspace}

\providecommand{\UfourS}{\mbox{$\Upsilon(4S)$}}
\newcommand{\etal}{{\em et al.}}
\newcommand{\jprlBase}       {Phys.\ Rev.\ Lett.\xspace}
\newcommand{\jprl}      [1]  {\jprlBase\ {\bf #1}}
\newcommand{\jprBase}        {Phys.\ Rev.\xspace}
\newcommand{\jprd}      [1]  {\jprBase\ D~{\bf #1}}
\newcommand{\jplBase}        {Phys.\ Lett.\xspace}
\newcommand{\plb}       [1]  {\jplBase\ B~{\bf #1}}
\newcommand{\nimBaseA}       {Nucl.\ Instr.\ Meth.\xspace}
\newcommand{\nima}      [1]  {\nimBaseA~A~{\bf #1}}
\newcommand{\epjBase}       {Eur.\ Phys.\ J.\xspace}
\newcommand{\epjc}      [1]  {\epjBase\ C~{\bf #1}}

\newcommand{\gevcc}{\ensuremath{{\mathrm{\,Ge\kern -0.1em V\!/}c^2}}\xspace}
\newcommand{\gevccq}{\ensuremath{{\mathrm{\,Ge\kern -0.1em V^2\!/}c^4}}\xspace}

\newcommand{\aunopm}{\mbox{$a_1(1260)^{\pm}  $}}
\newcommand{\pimp}{\mbox{$\pi^{\mp}  $}}
\newcommand{\appi} {\ensuremath{\aunopm\ \pimp\ }\xspace}

\begin{document}
{\pagestyle{empty}

\begin{flushright}
SLAC-PUB-\SLACPubNumber \\
\babar-PROC-\BABARPubYear/\BABARProcNumber \\
January, 2009 \\
\end{flushright}

\par\vskip 4cm

\begin{center}
\Large \bf Rare $B$ Decays at \babar
\end{center}
\bigskip

\begin{center}
\large 
F. Palombo\\
Universit\`a di Milano, Dipartimento di Fisica and INFN, I-20133
Milano, Italy \\
(for the \lbabar\ Collaboration)
\end{center}
\bigskip \bigskip

\begin{center}
\large \bf Abstract
\end{center}
I present some of the most recent \babar\  measurements for
rare $B$ decays. These include  rate asymmetries in the $B$ decays to $K^{(*)} l^+ l^-$ and
 $K^{+} \pi^{-}$ and  branching fractions in the $B$ decays   to  $
 l^+ \nu_l$,  $K_1(1270)^+ \pi^-$ and  $K_1(1400)^+ \pi^-$. 
I also
report a search for the $B^+$ decay to $K^0_S K^0_S \pi^+$.   

\vfill
\begin{center}
Contributed to the Proceedings of the 18$^{th}$ International 
Conference on Particles and Nuclei, \\
11/09/2008---11/14/2008, Eilat, Israel
\end{center}

\vspace{1.0cm}
\begin{center}
{\em Stanford Linear Accelerator Center, Stanford University, 
Stanford, CA 94309} \\ \vspace{0.1cm}\hrule\vspace{0.1cm}
Work supported in part by Department of Energy contract DE-AC02-76SF00515.
\end{center}
\newpage%

\section{Introduction}
The study of rare $B$ decays plays a central  role in the physics
program of the \babar\  \cite{babar} and Belle \cite{belle} 
experiments at the B-factories. Rare $B$ decays  
in fact provide sensitivity to the Standard Model (SM)
parameters. They are also powerful probes for   the presence of New 
Physics (NP).

\section{Results}
I present preliminary results of some of the latest  \babar\
 analyses  of electroweak, pure leptonic, and hadronic
penguin B decays. Most of these improved measurements are based on 
 the final \babar\ dataset of about $465 \times  10^6 $  \BBbar\  pairs taken at the  \UfourS.

\subsection{ Electroweak $B$ Decays  to $K^{(*)} l^+ l^-$} 
Electroweak decays $B \ra K^{(*)} l^+ l^-$ $(l=e,\mu)$ are
mediated by flavor-changing neutral current processes, forbidden
at tree level in the SM. These decays may proceed
 through a $\gamma$/Z loop (penguin) or a $W^- W^+$ box 
diagram. Contributions from NP
may enter the loop and box diagrams 
at  the same order as the SM ones and affect
both rates and kinematic distributions \cite{Ali}.
In  exclusive $B$ decay modes sensitivity to NP in the rates is
limited by hadronic uncertainties. For this reason it is preferable to
search for NP in rate asymmetries where  part of these uncertainties
cancels.  

\babar\  recently updated \cite{Kstll} measurements 
of  direct \CP asymmetry  $\calA^{K^{*}}_{\CP}$,
lepton flavor ratio \RK, and  \CP-averaged isospin asymmetry \calAI. $\calA^{K^{*}}_{\CP}$ and    \RK\ , expected in SM at a level of $10^{-4}$
and 1  respectively, could be  enhanced by NP contributions \cite{NP}.
Results for both  $\calA^{K^{*}}_{\CP}$ and \RK\ are in agreement  with SM 
expectations.  
\calAI\ is expected at a level of 0.01 in SM 
\cite{Feldmann}.  In combined fit to $K l^+ l^-$ and $K^* l^+ l^-$ in
low dilepton mass squared   region $0.1 < m^2_{ll}< 7.02$ \gevccq,
the fit  result is 
$\calAI= -0.64^{+0.15}_{-0.14} \pm 0.03$. It is  3.9 standard
deviations ($\sigma$)
(systematic uncertainties  included) from the null asymmetry.  
This effect is not seen in high $m^2_{ll}$ region ($>10.24 $ \gevccq).

\subsection{Pure Leptonic $B^+$ Decays to $l^+ \nu_l$}
The purely leptonic $B$ decays to $l \nu_l$ ($l=\tau ,\mu ,e$)
proceed in SM through $W$ boson annihilation with a branching
fraction (BF):  ${\cal B}(\Bp\ra l^+ \nu) = \frac{G^2_F m_B}{8 \pi} m_l^2 ( 1-
\frac{m^2_l}{m^2_B} )^2 f^2_B |V_{ub}|^2 \tau_{\Bp} $  \cite{lnu}.
The SM estimate of the BF for  $B \ra  \tau {\nu_\tau}$   is of the order of
$10^{-4}$. BFs for   modes with $\mu {\nu_\mu}$ and $e {\nu_e}$ are helicity
suppressed ($\sim  m^2_l$) by factors 225 and $10^{7}$,
respectively. However
contributions from NP scenarios \cite{Hou} may  enhance these SM expectations. 
\babar\  studied  these $B$ decays, 
searching signals with a  semi-leptonic tagging method \cite{lnu}.
The measured BF $\calB (\Bp\ \ra \tau^+ \nu_{\tau} )$ is 
$(1.8 \pm 0.8 \pm 0.1) \times 10^{-4} $  with a significance of  
$2.4 \, \sigma$. The upper limit (UL) at 90\% confidence
level (CL) is $3.2 \times 10^{-4}$.  
Combining this with  the other  \babar\  BF measurement of 
the same $B$ decay using an hadronic tag
\cite{lnuHTag}, we have $\calB( \Bp \ra \tau^+ \nu_{\tau} ) = (1.8 \pm 0.6) \times
10^{-4} $ with a significance of $3.2 \, \sigma$.

ULs at 90 \% CL are set for the other two decay modes: 
$\calB(\Bp\ra \mu^+ \nu_{\mu} ) < 11 \times 10^{-6}$ and 
$\calB(B^+\ra e^+ \nu_{e} ) < 7.7 \times 10^{-6}$. \babar\ also 
searched for the  $\Bp\ \ra \mu^+ \nu_{\mu}$ mode using an inclusive tag
and set an UL of $1.3 \times 10^{-6}$ at 90 \% CL \cite{InclTag}.
 
\subsection{Hadronic-Penguin $B$ Decays to K $\pi$, $K_1 \pi$ and \K0S
\K0S $\pi$}

At the B-factories direct \CP asymmetry has been observed in neutral $B$
decays to $\pi^+ \pi^-$ and $K^+ \pi^-$. \babar\  has recently
updated these measurements \cite{Kpi}. The measured direct \CP asymmetry
$\calA_{K^{\pm} \pi^{\mp}}$ between $\Bzb \ra K^- \pi^+ $ and $\Bz \ra
K^+ \pi^- $ is  $-0.107 \pm 0.016^{+0.006}_{-0.004}$ with a
significance  of $6.1 \, \sigma$. \babar\ also
measured the direct \CP asymmetry  $\calA_{K^{\pm}\pi^0}
=0.030 \pm 0.039 \pm 0.010$ \cite{Kpi0}.  The difference of these two
asymmetries, naively expected zero in SM, 
is: ${\Delta \calA} = \calA_{K^{\pm}\pi^0} - \calA_{K^{\pm} \pi^{\mp}} =0.137
\pm 0.044 $ with a significance at a level of $3.1 \, \sigma$.
This result is consistent with Belle measurement \cite{Nature}.
Using the world average of existing (\babar, Belle, CDF, CLEO)
results  \cite{HFAG} ,  we have $\Delta \calA =
0.148 \pm 0.028$. with a significance at the level of  $5.3 \, \sigma.$   
Several approaches try to explain the large difference of 
$\calA_{K^{\pm}\pi^0}$ and $ \calA_{K^{\pm} \pi^{\mp}}$ inside  the SM
as in \cite{G-R} or with contributions  from NP as in  \cite{WSH}. 

\babar\ measured \CP violating parameters in  
$\Bz(\Bzb) \ra \appi\ $~\cite{A1Pi} for  the extraction of the
$\alpha$ angle of the unitary triangle. 
To determine by means of SU(3) flavor-symmetry \cite{Zupan} the
uncertainties on  this $\alpha$ measurement due to
the presence of penguin contributions in the decay, \babar\ measured 
the BF of $B^0$ to $K_1(1270)^+ \pi^-$ 
and   $K_1(1400)^+ \pi^-$. Using a K-matrix formalism, a combined 
BF ${\cal B}(\Bz \ra K_1(1270)^+ \pi^- + K_1(1400)^+ \pi^-)$ 
 = $(31.0 \pm 2.7 \pm 6.9) \times 10^{-6}$ is  obtained with a
 significance $> 5.1 \, \sigma$ \cite{Simone}.

The $B^+$ decay to $ \K0S \K0S \pi^+$ is suppressed in SM with an
expected BF of the order $10^{-6}$ \cite{KsKspi}. An UL of $3.2 \times
10^{-6}$ at 90 \% CL has been set by Belle \cite{KsKspiBelle}.
The \babar\  analysis  has been done inclusively, incorporating
resonant and nonresonant intermediate states. 
No significant signal has been observed, setting an UL of $5.1 \times
10^{-7}$ at 90 \% CL \cite{KsKspiBaBar}.
In the $B^+$  decay to $K^+ K^- \pi^+$ \babar\ has seen a peak
($f_x(1500$)) at 1.5 \gevcc\ in the $K^+ K^-$ invariant mass \cite{Babar2}.
Because there is no evidence of this peak  in \K0S \K0S invariant
mass, models  with
even spin of $f_x(1500)$ and decays with isospin symmetry are disfavored.

\section{Summary and Conclusions}
I have presented a selection  from  the most  recent  \babar\ results 
on rare $B$ decays. No significant deviations from the SM expectations
have been seen (including also all those measurements on rare $B$
decays non presented here). The disagreement with SM found in a few
cases may be explained 
either  with improved calculations within the SM or with NP contributions.
In conclusion, we need much more precise measurements  to improve our
sensitivity to SM deviations and to NP effects. This task,
complementary  to LHC  NP search, can
be accomplished at LHCb \cite{LHCb} and at  a super B-factory \cite{superB}.   

\end{document}